\documentclass[twocolumn,a4paper,10pt,]{article}
\usepackage[top=0.85in,left=0.75in,footskip=0.75in]{geometry}

\usepackage{amsmath,amssymb}

\usepackage{changepage}

\usepackage[utf8x]{inputenc}
\usepackage{authblk}

\usepackage{textcomp,marvosym}

\usepackage{cite}

\usepackage{nameref,hyperref}

\usepackage{adjustbox}

\usepackage{microtype}
\DisableLigatures[f]{encoding = *, family = * }

\usepackage[table]{xcolor}

\usepackage{array}

\newcolumntype{+}{!{\vrule width 2pt}}

\providecommand{\keywords}[1]
{
  \small	
  \textbf{\textit{Keywords}} #1
}

\usepackage[aboveskip=1pt,labelfont=bf,labelsep=period,justification=raggedright,singlelinecheck=off]{caption}

\makeatletter
\renewcommand{\@biblabel}[1]{\quad#1.}
\makeatother

\usepackage{fancyhdr,graphicx} 
\usepackage{epstopdf}

\title{Global health science leverages established collaboration network to fight COVID-19}

\author{Stefano Bianchini}
\author{Moritz M\"uller} 
\author{Pierre Pelletier}
\author{Kevin Wirtz}

\affil[]{BETA, Universit\'{e} de Strasbourg, France}
\affil[ ]{\textit {\{s.bianchini,mueller,p.pelletier,kevin.wirtz\}@unistra.fr}}

\date{\today}

\begin{document}

\maketitle


\section*{Abstract}
{\small \bf How has the science system reacted to the early stages of the COVID-19 pandemic? Here we compare the (growing) international network for coronavirus research with the broader international health science network. Our findings show that, before the outbreak, coronavirus research realized a relatively small and rather peculiar niche within the global health sciences. As a response to the pandemic, the international network for coronavirus research expanded rapidly along the hierarchical structure laid out by the global health science network. Thus, in face of the crisis, the global health science system proved to be structurally stable yet versatile in research. The observed versatility supports optimistic views on the role of science in meeting future challenges. However, the stability of the global core-periphery structure may be worrying, because it reduces learning opportunities and social capital of scientifically peripheral countries --- not only during this pandemic but also in its ``normal'' mode of operation.}\\

\keywords{COVID-19 $|$ Scientific Networks $|$ International Collaboration $|$ Health Sciences} 

\section*{Introduction}

International scientific collaboration is on the rise since the early 1980s \cite{Adams2013}. The phenomenon is one aspect of globalization in science. International collaboration is observed 
in particular among productive researchers from top-tier universities located in advanced national scientific systems \cite{PanEtAl2012,JonesEtAl2008}. The gain is (more) excellent research \cite{Adams2013,PanEtAl2012}. The tendency of `excellence-attracting-excellence', however, entails the risk of increasing stratification not only within but also between national science systems \cite{JonesEtAl2008,HorlingsVandenbesselaar2011}. In order to catch-up scientifically, or at least not to fall behind, being well connected to the global knowledge flows has become a science policy imperative in most countries. 

The paper at hand treats the outbreak of the novel coronavirus Sars-CoV-2 in January 2020 as an exogenous shock to the international health science system. Our main interest is in the structural effects of the shock on the international health science network. Recent empirical studies have shown that the scientific contribution to coronavirus related research from individual countries has been very uneven; often framing it as a scientific race \cite{AvivreuvenRosenfeld2020,RadanlievEtAl2020}. \cite{FryEtAl2020} investigate the international coronavirus collaboration network, and find that it has become more `elitist' with the pandemic. 

Our empirical analysis adds the insight that the contribution of countries to coronavirus research is closely related to their contribution in the broader domain of health sciences, and that the structure of the international coronavirus research network rapidly converged to the structure of the global, international health science network. Before we discuss the implications of this finding, let us first turn to the empirical analysis.

\section*{Data}

We proxy scientific activity in the health sciences through peer-reviewed articles in journals indexed by MEDLINE. The restriction to MEDLINE indexed journals ensures that papers in the sample fall into our scope of biomedical research and are of (minimum) scientific quality. Coronavirus related papers are identified through a text search query suggested by PubMed Central Europe on the papers' title, abstract, and MESH terms.

\begin{figure*}[!ht]
\includegraphics[width=\textwidth]{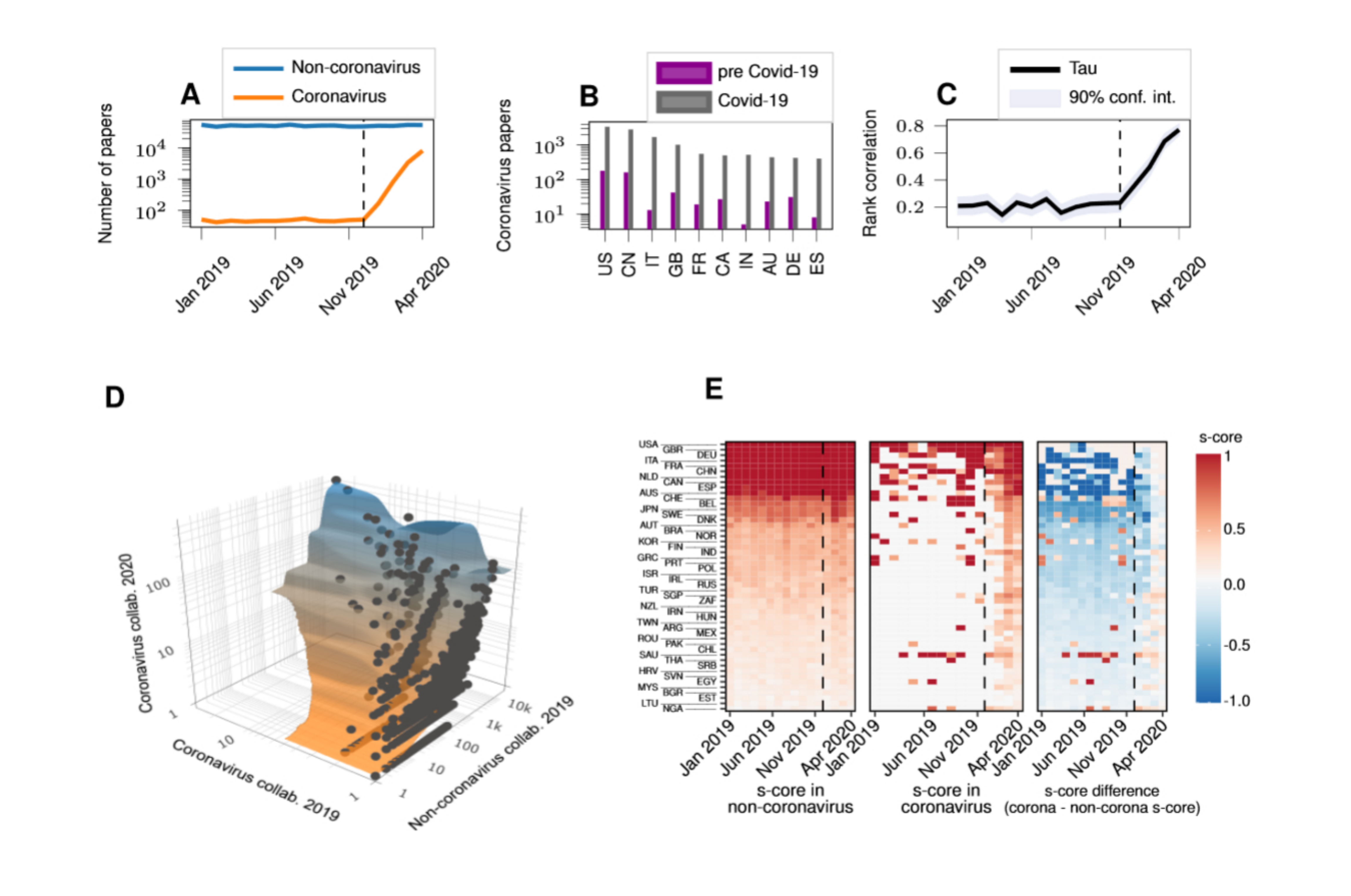} %
\caption{{\bf Countries take on the same role in coronavirus research as in the global health sciences.}
{\small (A) Coronavirus and non-coronavirus papers by month. (B) Top 10 countries in coronavirus-related research during COVID-19. (C) Correlation of country rankings by coronavirus and non-coronavirus research by month. (D) Surface plot of a local regression of (log of) joint coronavirus papers during COVID-19 on (log of) joint coronavirus and non-coronavirus papers pre-COVID-19. (E) Country centrality based on s-core decomposition of the coronavirus and non-coronavirus network by month.}}
\label{figure1} 
\end{figure*}

The analysis is based on the papers' submission dates to stay close to the actual research activity. Our working sample includes papers submitted in the pre-COVID-19 period (Jan.--Dec.2019), as well as in the early phase of COVID-19 (Jan.--Apr.2020). In detail, we downloaded all papers appearing in MEDLINE journals from the PubMed database as of December 2020. Due to the time lag from submission to acceptance, the number of paper submissions in our sample starts to drop in May; a data artifact that may bias statistics. Therefore, we end the analysis period in April 2020. 

Our final working sample consists of 837,427 papers.  Distinguishing coronavirus related research from non-coronavirus related research, and pre-COVID-19 period (Jan.--Dec.2019) from COVID-19 period (Jan.--Apr.2020) yields four categories: 614,141 non-coronavirus, pre-COVID-19 papers, 571 coronavirus, pre-COVID-19 papers, 210,171 non-coronavirus, COVID-19 papers, and 12,544 coronavirus, COVID-19 papers.\footnote{All data and scripts are available on gitHub \url{https://github.com/P-Pelletier/Global-health-sciences-response-to-COVID-19}.}

\section*{Results}


We first count papers per month (Fig~\ref{figure1}A). In the pre-COVID-19 period, coronavirus research output is relatively stable, at roughly 50 papers per month. Starting with the January 2020 outbreak, coronavirus research grows exponentially up to 8,159 submissions in April 2020. Other research output is stable throughout, at about 51,000 papers, and even increases slightly with the pandemic. Apparently, many (male) researchers took advantage of the lockdown period to finish off research that piled up already before COVID-19 \cite{BellFong2021}. Potentially negative effects of the pandemic due to frictions in the research machinery, or crowding out of non-coronavirus research are not yet visible in this early period.

\subsection*{National scientific production}

Next, consider the contribution of individual countries to coronavirus research. We employ a full-count assignment scheme -- i.e., each paper with at least one affiliation in a given country counts fully (one) for that country. The distribution is highly skewed: the 10 most prolific countries generate 70 percent of coronavirus research during COVID-19; ranging from the US signing 2,686 to Spain with 381 papers (Fig~\ref{figure1}B). All these countries are big players in the health sciences, but not all bring in a strong track record in coronavirus research. 

So, how important was coronavirus-specific research capacity compared to general health science capacity in the early response? Simple linear regressions of coronavirus papers in Jan.-- Apr.2020 on pre-pandemic coronavirus and other paper counts provide some indication (Table~\ref{table1}). All variables in the regression are transformed into logs, and normalized to zero mean and unit variance to facilitate direct interpretation of coefficient estimates. We find that pre-pandemic coronavirus research is highly (and significantly) correlated with coronavirus research in January 2020, while other research is not. However, this pattern reverses within the next three months when coronavirus research takes off. Note also that variations in the outcome variable are increasingly well explained, with $R^2$ from 0.7 in January to 0.9 in April 2020; mostly due to prior non-coronavirus research. 

\begin{table}[!h]
\caption{Coronavirus papers (2020) on prior papers (2019).}
\begin{adjustbox}{width=\columnwidth,center}
\begin{tabular}{lccccc}
  & {\bf Jan. `20} & {\bf Feb. `20} & {\bf Mar. `20} & {\bf Apr. `20} & {\bf Total} \\ 
   \hline
  Coronav. `19 & 0.859 & 0.727 & 0.431 & 0.262 & 0.231 \\ 
    & (0.131) & (0.077) & (0.044) & (0.036) & (0.031) \\ 
    \hline
  Others `19 & 0.008 & 0.198 & 0.564 & 0.732 & 0.765 \\ 
   & (0.067) & (0.055) & (0.044) & (0.039) & (0.035) \\ 
  \hline
  $R^2$ & 0.748 & 0.788 & 0.876 & 0.898 & 0.910 \\ 
  \hline
 \end{tabular}
\label{table1}
\end{adjustbox}
\begin{flushleft} \footnotesize Notes: 200 observations. All variables in logs, with zero mean and unit variance. Standard errors in parenthesis.
\end{flushleft}
\end{table}

By the same token, countries take rapidly very similar positions in rankings on coronavirus papers as they do in rankings on other health papers (see Fig~\ref{figure1}C). We calculate the rank correlation coefficient $\tau_X$ of \cite{EmondMason2002}. It is similar to Kendall's $\tau$, but handles ties the same as dominant relationships (i.e. entering 1 and not 0 in the dominance matrix). This is favorable in case of many ties in the rankings, as we have in corona pre-Covid-19 research, but does not really change the results. The 90 percent confidence interval around $\tau_X$ has been obtained through a traditional jackknife, or leave-one-out, approach (see e.g. \cite{AbdiWilliams2010}). Fig~\ref{figure1}C shows the result. Until the outbreak in January 2020 (vertical dashed line in Fig~\ref{figure1}C) rank correlations are rather low at around 0.2. After the outbreak, the (monthly) scientific output of countries in corona aligns with non-corona research output until a (high) correlation of 0.8 in April 2020.

We summarize the first part of the analysis. Before the pandemic, leading countries in the health sciences have not necessarily led coronavirus research. Within a few months after the COVID-19 shock, leading countries in the health sciences also led coronavirus research. The second part of the analysis establishes the same dynamic for the international collaboration networks.

\subsection*{International scientific collaboration}

\begin{figure*}[!h]
\includegraphics[width=\textwidth]{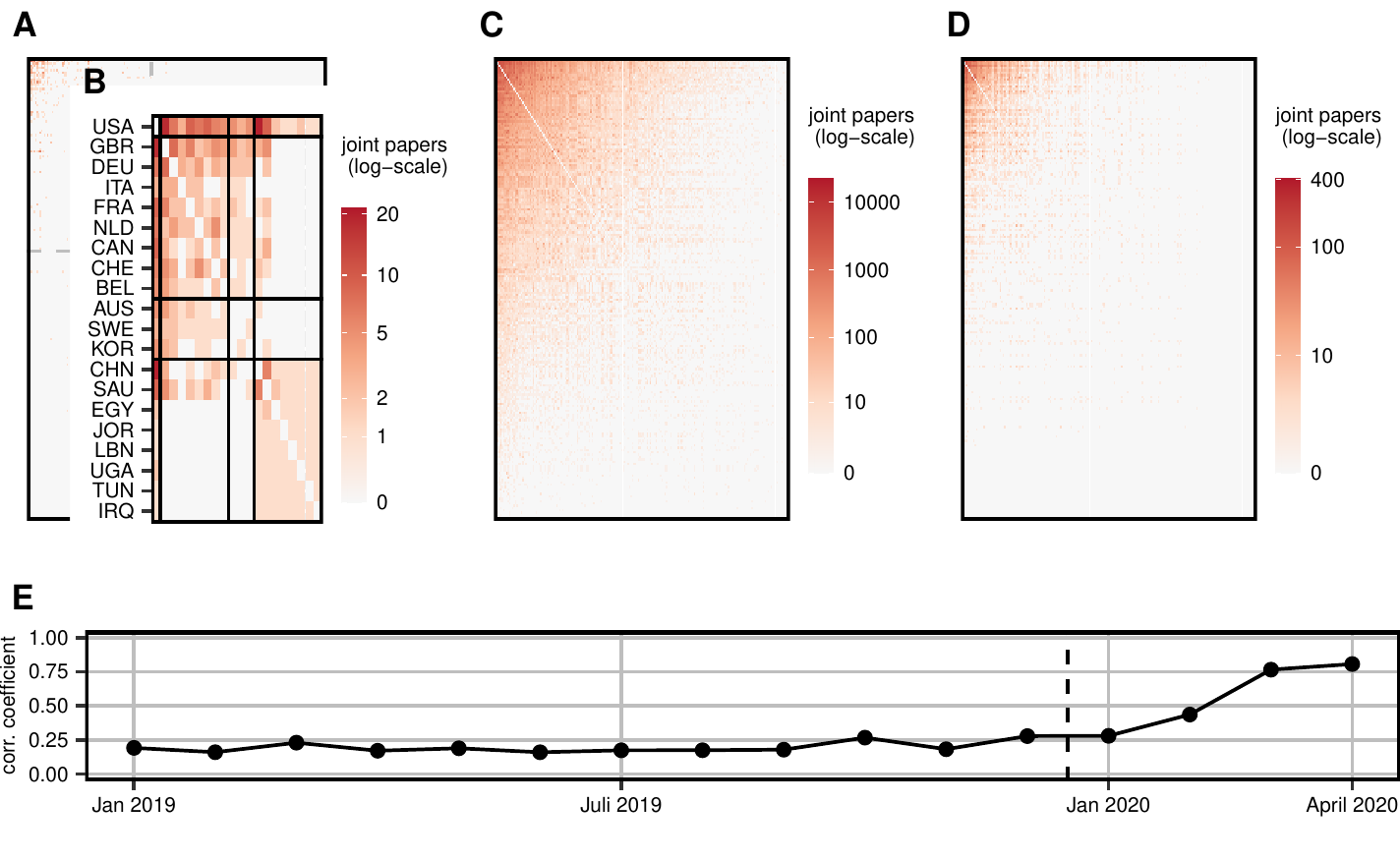}
\caption{{\bf Coronavirus research network converges to global health science network.}
{\small A) to D) are (accumulated) adjacency matrices of A) and B) coronavirus, pre-COVID-19 network, C) non-coronavirus, pre-COVID-19 network, D) coronavirus, COVID-19 network. In A), C) and D) countries ordered by eigenvector centrality in non-coronavirus, pre-COVID-19 network (C). In B) countries ordered through generalized block modeling. E) correlation coefficient of the (monthly) coronavirus and non-coronavirus networks. All correlations obtained are highly significant in QAP test (p$<$0.001).}}
\label{figure2}
\end{figure*}

We construct international scientific collaboration networks based on co-authorship of papers in our sample. A node corresponds to a country. Edge weights correspond to the number of joint papers (full-accounting scheme).

First, we consider link formation. Are prior collaborations on coronavirus research (i.e. same topic), or any prior ties (i.e. different topics) replicated for coronavirus research during the pandemic? Fig~\ref{figure1}D provides some indication. The surface plot is obtained from a local regression with least-squares cross-validated bandwidths for the local constant estimator. It shows the expected (log of) joint coronavirus papers during COVID-19, conditional on (log of) joint coronavirus and non-coronavirus papers pre-COVID-19. Looking at Fig~\ref{figure1}D, we first note that most country-pairs (dots) had no pre-COVID-19 joint coronavirus research. Their number of joint coronavirus papers during COVID-19 increases with the number of other joint papers before the pandemic. As we increase from zero prior coronavirus papers, (expected) joint coronavirus papers during COVID-19 (the surface in Fig~\ref{figure1}D) also increases. Yet, it is evident that bi-national collaboration on coronavirus related research after the shock largely reflects bi-national collaboration on non-coronavirus research before the shock. 

Consequently, countries' network centrality in the coronavirus research network aligns with their centrality in the overall health science network. We capture network centrality through (normalized) s-core decomposition \cite{EidsaaAlmaas2013}. The s-core ranges from 0 for isolates in the network, to 1 for (highest) core members. Fig~\ref{figure1}E provides the monthly s-cores on the non-coronavirus network (left panel), coronavirus network (middle panel), and the difference of s-cores in coronavirus and non-coronavirus networks (right panel). The figure shows the 60 most central countries in the non-coronavirus network and applies that same ordering across all three panels. The remaining countries are highly peripheral in the considered networks. The left panel shows that the global network hierarchy is very stable. The core is formed by (mostly large) countries of the global north, China being the exception. Centrality in the coronavirus network is more dynamic (middle panel). Pre-COVID-19, most countries are not involved in coronavirus related collaborations and, hence, in the extreme periphery (white). The core of the coronavirus network includes only a few countries leading other health sciences. Saudi Arabia stands out as it is part of the core in the coronavirus network, but peripheral in the overall health science network. (Variations in core membership over time may be explained by lower research activity overall which leads to more erratic signals.) After the shock, the structure of the coronavirus network shifts rapidly towards the hierarchy in health science at large. This is easily seen in the right panel that shows the difference between the s-core centrality in the coronavirus and non-coronavirus network. Prior to the shock, s-core differences range from -1 in dark blue (for countries at the extreme periphery in the coronavirus network and in the core in the other network), over 0 in white (same s-core in both networks), up to 1 in red (for countries in the coronavirus network core and peripheral in the non-coronavirus network). After the shock, the global core rapidly takes its role in coronavirus related research, and so does the global periphery (all countries appear in light colors with an s-core difference of around zero in April 2020).

Fig~\ref{figure2} pictures the networks in form of adjacency matrices. In order to facilitate a comparison across networks, adjacency matrices of the pre-pandemic coronavirus network (A), pre-pandemic non-coronavirus network (C), and pandemic coronavirus network (D) are all ordered by eigenvector centrality in (C). 

The pre-pandemic coronavirus network (A) is relatively sparse and best described by a block model (B): A regional middle east community and a community of mostly developed countries, connected through USA, China, and Saudi Arabia. The block model is obtained by minimizing the absolute difference of the number of papers in logs over blocks. This finding is robust to alternative algorithms. Essentially the same (community) structure transpires from hierarchical community detection on the weighted pre-pandemic coronavirus network using the OSLOM \cite{LancichinettiEtAl2011} as well as the LOUVAIN \cite{BlondelEtAl2008} algorithm. 

In contrast, we find that the pre-pandemic, non-coronavirus network (C) corresponds to a nested-split graph. A nested-split graph is a specific type of a hierarchical network, in which the most central node connects to all other nodes, and less central nodes connect to subsets of alters of more central nodes. Nested-split graphs emerge in network games where payoffs are strategic complements in effort levels \cite{KoenigEtAl2014}, which is a reasonable assumption for science networks. The coronavirus network after the shock (D) closely resembles the non-coronavirus network before the shock (C). The correlation coefficient (E) of the (monthly) coronavirus and non-coronavirus adjacency matrices (in logs) confirms that convergence. All correlations are highly significant; based on a QAP test that creates a null distribution through re-labeling of nodes while maintaining the structure of the networks \cite{Krackhardt1987}.  

\section*{Conclusion}

More elitist science in the COVID-19 era? Novel ways of organizing science? Not really. The structural shift towards a highly hierarchical system in coronavirus related research mostly picks up established structures of the broader, globalized health sciences. Thus, we conjecture that coronavirus research during the pandemic will further sustain, not break with, long-term trends in international collaborations. The pandemic feeds into an ongoing global stratification in science that reduces systematically learning opportunities and social capital of scientifically peripheral countries. Policy should therefore aim at a more inclusive science landscape --- not only during crises, but even more in its `normal mode of operation'. 

\section*{Acknowledgments}
The research leading to the results of this paper has received financial support from the CNRS through
the MITI interdisciplinary program Enjeux scientifiques et sociaux de l'intelligence artificielle-AAP 2020 [reference: ARISE].

\clearpage
\newpage


%
%
%

\bibliography{}


\end{document}